\documentclass[a4paper,12pt]{article}

\pdfoutput=1
\usepackage{amsmath}
\usepackage{amssymb}
\usepackage{amsfonts}
\usepackage{mathrsfs}
\usepackage{bbm}
\usepackage{graphicx,subfigure}

\usepackage{bookmark}

\usepackage{cite}

\usepackage{calc}
\usepackage{hyperref}
\usepackage{array}

\usepackage{ulem}

\usepackage{multirow}
\usepackage{color}
\usepackage{xcolor,colortbl}

\usepackage{psfrag}
\usepackage{pstricks}
\usepackage{epsfig}

\allowdisplaybreaks

\newlength{\dinwidth}
\newlength{\dinmargin}
\definecolor{nicered}{rgb}{1.0,0.0,0.2}

\definecolor{color1}{rgb}{0.9,.4,.2}

\definecolor{color2}{rgb}{0.3,.6,.7}

\definecolor{color3}{rgb}{0.7,.2,.7}

\usepackage{color}

\begin{document}

\title{
\vspace*{-0.5cm}
\bf \Large
Study of  $a_{0}^{0}(980)-f_0(980)$ mixing from $\mathbf{\boldsymbol{a_0(1450)\rightarrow a_{0}^{0}(980)f_0(500)\rightarrow{\pi}^+{\pi}^-f_0(500)}}$}

\author{Xiao-Dong Cheng$^{1}$\footnote{chengxd@mails.ccnu.edu.cn},  Ru-Min Wang$^{2}$, Yuan-Guo Xu$^{2}$\\
\\
{$^1$\small College of Physics and Electronic Engineering,}\\[-0.2cm]
{    \small Xinyang Normal University, Xinyang 464000, People's Republic of China}\\[-0.1cm]
{$^2$\small  College of Physics and Communication Electronics,}\\[-0.2cm]
{    \small  JiangXi Normal University, NanChang 330022, People's Republic of China}\\[-0.1cm]}

\date{}
\maketitle
\bigskip\bigskip
\maketitle
\vspace{-1.2cm}

\begin{abstract}
{\noindent}The $a_0^0(980)-f_0(980)$ mixing is one of the most potential tools to learn about the nature of $a_0^0(980)$ and $f_0(980)$. Using the $f_0(980)$-$a_0^0(980)$ mixing intensity $\xi_{af}$ measured recently at BESIII, we calculate the the branching ratio of the the isospin violation decay $J/\psi \rightarrow\gamma\eta_c \rightarrow \gamma \pi^0 a_0^0(1450)\rightarrow \gamma \pi^0 a_0^0(980)f_0(500)\rightarrow \gamma \pi^0 f_0(980) f_0(500) \rightarrow \gamma \pi^0 \pi^+\pi^- \pi^+\pi^-$. The value of the branching ratio is found to be $O(10^{-6})$,  which can be observed with $10^{10}$ $J/\psi$ events collected at BESIII. The narrow peak from the $f_0(980)$-$a_0^0(980)$ mixing in the $\pi^+\pi^-$ mass square spectrum can also be observed. In addition, we study the non-resonant decay $a_0^0(1450)\rightarrow  f_0(980) \pi^+\pi^-(\text{non-resonant})$, which is dominated by the $a_0^0(980)$-$f_{0}(980)$ mixing. We find that the non-resonant decay $a_0^0(1450)\rightarrow  f_0(980) \pi^+\pi^-$ and the decay $a_0^0(1450)\rightarrow  f_0(980) f_0(500)$ can be combined to measure the mixing intensity $\xi_{af}$ in experiment. These decays are the perfect complement to the decay $\chi_{c1}\rightarrow f_{0}(980)\pi^{0}\to\pi^{+}\pi^{-}\pi^{0}$ which had been observed at BESIII, the observations of them will make the measurement of the mixing intensity $\xi_{af}$ more precisely.
\end{abstract}
\newpage

\section{Introduction}
\label{sec:intro}
The inner structure of the light scalar mesons such as $a_0^0(980)$ and $f_0(980)$ has been studied for over thirty years, and now it is still a hot topic in particle physics. There are several proposals for the inner structure of the light scalar mesons, such as $q \bar{q}$ states, glueball, hybrid states, molecule states, tetra-quark states and the superpositions of these contents~\cite{Cheng:2005nb, Weinstein:1982gc, Weinstein:1983gd, Weinstein:1990gu, Jaffe,Kim:2017yur,Abdel-Rehim:2014zwa, Amsler:1995td, Amsler:1995tu, Amsler:2002ey,Gorishnii:1983zi}. However, there is still no general agreement on the inner structure of $a_0^0(980)$ and $f_0(980)$, due to the absence of convincing evidence.

The $a_0^0(980)-f_0(980)$ mixing, which was first suggested theoretically in Ref.~\cite{Achasov:1979xc}, is one of the most potential tools to learn about the nature of $a_0^0(980)$ and $f_0(980)$, and therefore has been studied extensively in various processes~\cite{Achasov:1981de,Achasov:1996qn,Krehl:1996rk,Kerbikov:2000pu,Close:2000ah,Kudryavtsev:2001ee,Grishina:2001zj,Close:2001ay,Kudryavtsev:2002uu,
Kondratyuk:2002yf,Achasov:2002hg,Achasov:2003se,Grishina:2004rd,Achasov:2004ur,Wu:2007jh,Wu:2008hx,Hanhart:2007bd,Ablikim:2010aa,Wu:2011yx,Aceti:2012dj,Roca:2012cv,
Tarasov:2013yma,Sekihara:2014qxa,Aceti:2015zva,Achasov:2015uua,Wang:2016wpc,Achasov:2016wll,Achasov:2017edm,Achasov:2017zhu,Sakai:2017iqs,Bayar:2017pzq,Achasov:2017ncx,
Liang:2017ijf,Achasov:2019ywj,Achasov:2019zrc,Liang:2019jtr,Molina:2019udw,Achasov:2020fee,Achasov:2020qfx}.In February 2018, The BESIII Collaboration studied the $a_0^0(980)-f_0(980)$ mixing with the decays of $J/\psi \rightarrow \phi f_0(980)\rightarrow \phi a_0^0(980)\rightarrow\phi\eta\pi^0$ and $\chi_{c1}\to a^{0}_{0}(980)\pi^{0}\to f_{0}(980)\pi^{0}\to\pi^{+}\pi^{-}\pi^{0}$, the signals of the $a_0^0(980)-f_0(980)$ mixing were observed with a statistical significance of larger than $5\sigma$ for the first time. The values of the mixing intensities were mesaured ~\cite{Ablikim:2018pik}
\begin{align}\label{eq:epsilonfabesiii}
  \begin{split}
     \xi_{fa} &= (0.99 \pm 0.35) \times 10^{-2} ~~~ \text{(solution-1)} \\
     \xi_{fa} &= (0.41 \pm 0.25) \times 10^{-2} ~~~ \text{(solution-2)}~.
   \end{split}
\end{align}
and
\begin{align}\label{eq:epsilonafbesiii}
  \begin{split}
     \xi_{af} &= (0.40 \pm 0.17) \times 10^{-2} ~~~ .
   \end{split}
\end{align}
here, the mixing intensities $\xi_{af}$ and $\xi_{fa}$ are defined as
\begin{align}\label{eq:epsilonfadefine}
\xi_{fa}=\frac{{\mathcal B}(J/\psi \rightarrow \phi f_0(980)\rightarrow \phi a_0^0(980)\rightarrow \phi \eta \pi^0)}{{\mathcal B}(J/\psi \rightarrow \phi f_0(980)\rightarrow \phi \pi^{+} \pi^{-})}\\
\xi_{af}=\frac{{\mathcal B}(\chi_{c1}\to a^{0}_{0}(980)\pi^{0}\to f_{0}(980)\pi^{0}\to\pi^{+}\pi^{-}\pi^{0})}{{\mathcal B}(\chi_{c1}\to a^{0}_{0}(980)\pi^{0}\to \eta\pi^{0}\pi^{0})}.
\end{align}
there are two solutions for the mixing intensity $\xi_{fa}$, the recent theoretical calculation prefer to the solution-1 result~\cite{Aliev:2018bln}. The result of $\xi_{af}$ suffers large uncertainty, and a question whether there would be a difference between the two mixing intensities $\xi_{af}$ and $\xi_{fa}$ may be raised, so more precise data and more reactions are needed in both experiment and theory.

The $a_0 (1450)$ resonance is a scalar-isovector meson and is assumed to be the conventional quark-antiquark structure based on the native quark model, the latest theoretical calculations~\cite{Mathur:2006bs,lee,Han:2013zg} also confirmed this conclusion. The BABAR collaboration performed a Dalitz plot analysis for the $\eta_c\rightarrow K^+ K^- \pi^0$ and $\eta_c\rightarrow K^+ K^- \eta$ decays and obtained the branch fraction of the $\eta_c\rightarrow a_0(1450) \pi^0\rightarrow K^+ K^- \pi^0$ decay relative to the $\eta_c\rightarrow K^+ K^- \pi^0$ mode~\cite{Lees:2014iua}
\begin{align}\label{eq:babaretcaazpiz}
\frac{{\mathcal B}(\eta_c \rightarrow a_0(1450)\pi^0)\cdot{\mathcal B}(a_0(1450) \rightarrow K^+ K^-)}{{\mathcal B}(\eta_c \rightarrow K^+ K^-\pi^0)}=(10.2\pm2.5)\times 10^{-2}.
\end{align}
By combining the recent data on the branching ratio of $\eta_c \rightarrow K^+ K^-\pi^0$ and $J/\psi \rightarrow \gamma \eta_c$ from the Particle Data Group~\cite{Tanabashi:2018oca}
\begin{align}
{\mathcal B}(J/\psi \rightarrow \gamma \eta_c)=(1.7\pm0.4)\times 10^{-2}\label{eq:pdgetctokkpiz1}~\\
{\mathcal B}(\eta_c \rightarrow  K^+ K^-\pi^0)=(3.65\pm0.25)\times 10^{-2}\label{eq:pdgetctokkpiz2}
\end{align}
and the value of the branching ratio of $a_0(1450) \rightarrow K^+ K^-$ in Ref.~\cite{Bugg:2008ig}
\begin{align}\label{eq:bugazztokpkm}
{\mathcal B}(a_0(1450) \rightarrow K^+ K^-)=(4.61\pm0.61)\times 10^{-2},
\end{align}
we find that the branching ratio of $J/\psi \rightarrow \gamma \eta_c\rightarrow \gamma a_0(1450)\pi^0$ can reach the order of $10^{-3}$. Based on the data samples of $10^{10}$ $J/\psi$ events collected with the BESIII experiment~\cite{Ablikim:2019hff,Asner:2008nq,Li:2016tlt,Bigi:2017eni}, about $10^{7}$ $a_0(1450)$ meson can be produced through decays $J/\psi \rightarrow \gamma \eta_c\rightarrow \gamma a_0(1450)\pi^0$, this large $a_0(1450)$ sample at BESIII will make it possible to investigate the properties of $a_0(1450)$ meson and study the related physics.

In this paper, we investigate the isospin breaking decay $\eta_c\rightarrow a_0(1450)\pi^0\rightarrow a_0^0(980) f_{0}(500)\pi^0\rightarrow \pi^+ \pi^- f_{0}(500)\pi^0$ produced via $J/\psi \rightarrow\gamma \eta_c$. We predict the branching ratio of this reaction by using the recent measurements at BESIII and calculate the distribution of the $\pi^+\pi^-$ mass square spectrum near the $K\bar{K}$ thresholds. We also discuss the $a_0(1450)\rightarrow f_0(980) \pi^+ \pi^-\text{(non-resonant)}\rightarrow \pi^+ \pi^- \pi^+ \pi^-$ decay process which is realized mainly via the $a_0^0(980)$-$f_0(980)$ mixing.
\section{the data on the decay}
\label{sec:dataonthedecay}
The $\eta_c\rightarrow \pi^0 a_0(1450)\rightarrow  \pi^0 a_0^0(980)f_{0}(500)\rightarrow\pi^0\pi^+ \pi^- f_{0}(500)$ decay violate the isospin symmetry, it can proceed via the $a_0^0(980)$-$f_0(980)$ mixing. In this process, the mixing intensity $\xi_{af}$ is given as
\begin{align}\label{eq:epsilonafazz}
\xi_{af}=\frac{{\mathcal B}( \eta_c\rightarrow \pi^0 a_0(1450)\rightarrow  \pi^0 a_0^0(980)f_{0}(500)\rightarrow \pi^0 f_0(980)f_{0}(500)\rightarrow \pi^0\pi^+ \pi^- \pi^+ \pi^- )}{{\mathcal B}( \eta_c\rightarrow \pi^0 a_0(1450)\rightarrow  \pi^0 a_0^0(980)f_{0}(500)\rightarrow  \pi^0\eta \pi^0 f_{0}(500)\rightarrow  \pi^0\eta \pi^0\pi^+ \pi^-)}.
\end{align}
here, ${\mathcal B}( \eta_c\rightarrow \pi^0 a_0(1450)\rightarrow  \pi^0 a_0^0(980)f_{0}(500)\rightarrow  \pi^0\eta \pi^0 f_{0}(500)\rightarrow  \pi^0\eta \pi^0\pi^+ \pi^-)$ is the branching ratio of the $\eta_c\rightarrow \pi^0 a_0(1450)\rightarrow  \pi^0 a_0^0(980)f_{0}(500)\rightarrow  \pi^0\eta \pi^0 f_{0}(500)\rightarrow  \pi^0\eta \pi^0\pi^+ \pi^-$ decay
\begin{align}\label{eq:bretaca98}
&{\mathcal B}( \eta_c\rightarrow \pi^0 a_0(1450)\rightarrow  \pi^0 a_0^0(980)f_{0}(500)\rightarrow  \pi^0\eta \pi^0 f_{0}(500)\rightarrow  \pi^0\eta \pi^0\pi^+ \pi^-)=\nonumber
\\&={\mathcal B}( \eta_c\rightarrow \pi^0 a_0(1450))\cdot{\mathcal B}(a_0(1450)\rightarrow  a_0^0(980)f_{0}(500))\cdot{\mathcal B}(a_0^0(980)\rightarrow  \eta \pi^0)\cdot{\mathcal B}(f_0(500)\rightarrow \pi^+ \pi^-).
\end{align}
By combining Eq.(\ref{eq:babaretcaazpiz}) and Eq.(\ref{eq:pdgetctokkpiz2}), one can obtain
\begin{align}\label{eq:bretacpiazzkk}
{\mathcal B}(\eta_c \rightarrow \pi^0 a_0(1450))\cdot{\mathcal B}(a_0(1450) \rightarrow K^+ K^-)=(3.72\pm0.96)\times 10^{-3}.
\end{align}
The ratio of the branching ratio of $a_0(1450) \rightarrow a_0^0(980)f_{0}(500)$ to $a_0(1450) \rightarrow K^+ K^-$ had been presented in Ref.~\cite{Bugg:2008ig}
\begin{align}\label{eq:ratiobrasigkk}
\frac{{\mathcal B}( a_0(1450) \rightarrow a_0^0(980)f_{0}(500))}{{\mathcal B}(a_0(1450) \rightarrow K^+ K^-)}=8.24\pm2.72,
\end{align}
so we can predict the branching ratio of the decay chain $\eta_c \rightarrow a_0(1450)\pi^0\rightarrow a_0^0(980)f_{0}(500)\pi^0$ as
\begin{align}\label{eq:brchainetacasi}
{\mathcal B}( \eta_c \rightarrow \pi^0 a_0(1450))\cdot{\mathcal B}(a_0(1450) \rightarrow a_0^0(980)f_{0}(500))=(3.06\pm1.28)\times 10^{-2}.
\end{align}
From Refs.~\cite{Cheng:2005nb,Cheng:2013fba,Ablikim:2018qzz}, we can obtain the branching ratios ${\mathcal B}(f_0(500)\rightarrow\pi^+ \pi^-)$, ${\mathcal B}( f_0 (980)\rightarrow \pi^+ \pi^-)$ and ${\mathcal B}(a_0^0(980)\rightarrow \eta \pi^0)$
\begin{align}
{\mathcal B}(f_0(500)\rightarrow\pi^+ \pi^-)=0.67,\label{eq:branchingratioofa9801}\\
{\mathcal B}( f_0 (980)\rightarrow \pi^+ \pi^-)=0.50^{+0.07}_{-0.09}\label{eq:branchingratioofa9802}\\
{\mathcal B}(a_0^0(980)\rightarrow \eta \pi^0)=0.845\pm 0.017\label{eq:branchingratioofa9803}.
\end{align}
Substituting Eq.(\ref{eq:branchingratioofa9801}), Eq.(\ref{eq:branchingratioofa9803}) and Eq.(\ref{eq:brchainetacasi}) into Eq.(\ref{eq:bretaca98}), one could predict the branching ratio of the decay chain $\eta_c \rightarrow\pi^0 a_0(1450)\rightarrow \pi^0 a_0^0(980)f_{0}(500)\rightarrow \pi^0\eta \pi^0 f_{0}(500)\rightarrow \pi^0\eta \pi^0\pi^+ \pi^-$ as
\begin{align}\label{eq:bretacasietapi}
{\mathcal B}( \eta_c\rightarrow \pi^0 a_0(1450)\rightarrow  \pi^0 a_0^0(980)f_{0}(500)\rightarrow  \pi^0\eta \pi^0 f_{0}(500)\rightarrow \pi^0\eta \pi^0\pi^+ \pi^-)=(1.74\pm0.72)\times 10^{-2}.
\end{align}
Combining this equation with Eq.(\ref{eq:epsilonafbesiii}) and using Eq.(\ref{eq:epsilonafazz}), we then obtain the branching ratio ${\mathcal B}( \eta_c\rightarrow \pi^0 a_0(1450)\rightarrow  \pi^0 a_0^0(980)f_{0}(500)\rightarrow \pi^0 f_0(980)f_{0}(500)\rightarrow \pi^0\pi^+ \pi^- \pi^+ \pi^-)$ as
$(0.70\pm0.41)\times 10^{-4}$. Adding this value and Eq.(\ref{eq:pdgetctokkpiz1}) together, we can readily obtain
\begin{align}\label{eq:bretacetapizsig}
&{\mathcal B}( J/\psi\rightarrow\gamma\eta_c\rightarrow\gamma \pi^0 a_0(1450)\rightarrow  \gamma\pi^0 a_0^0(980)f_{0}(500)\rightarrow \gamma\pi^0 f_0(980)f_{0}(500)&\rightarrow\gamma \pi^0\pi^+ \pi^- \pi^+ \pi^-)\nonumber
\\&=(1.19\pm0.75)\times 10^{-6}.
\end{align}
Obviously, this decay is hopefully to be marginally detected in the $e^+e^-$ colliders in view of the large database of BESIII.
\section{the mechanism responsible for the decay }
\begin{figure}[t]
\centering
\includegraphics[width=0.6\textwidth]{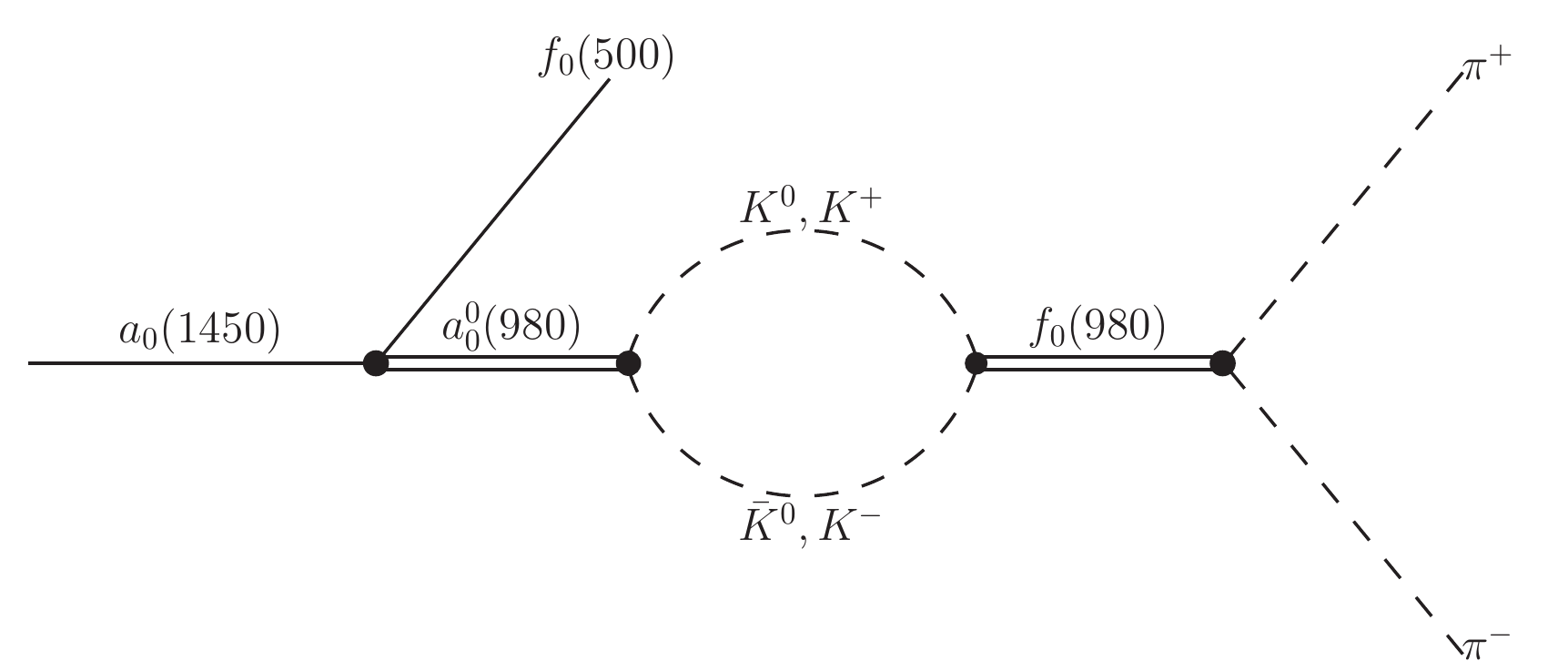}
\caption{\small Feynman diagram for the reaction $a_0(1450)\rightarrow a_0^0(980)f_{0}(500)\rightarrow f_0(980)f_{0}(500)\rightarrow\pi^+ \pi^- f_{0}(500)$.}
\label{aoffztosigmapizpif}
\end{figure}

As for the decay $a_0(1450)\rightarrow a_0^0(980)f_{0}(500)\rightarrow f_0(980)f_{0}(500)\rightarrow\pi^+ \pi^- f_{0}(500)$, the amplitude is proportional to the mixing of the $a_0^0(980)$ and $f_{0}(980)$ resonances, which is caused by the mass difference of $K^+K^-$ and $K^0 \bar{K}^0$ intermediate state. The diagram of $f_{0}(980)$ production in the $a_0(1450)\rightarrow a_0^0(980)f_{0}(500)\rightarrow f_0(980)f_{0}(500)\rightarrow\pi^+ \pi^- f_{0}(500)$ reaction is shown in Fig~\ref{aoffztosigmapizpif}, so the decay amplitude from this process can be written as
\begin{align}\label{eq:decayamplitudeofazzfa}
&{\mathcal M}(a_0(1450)\rightarrow a_0^0(980)f_{0}(500)\rightarrow f_0(980)f_{0}(500)\rightarrow\pi^+ \pi^- f_{0}(500)) =\nonumber\\
&~~={\mathcal M}_{a_0(1450)a_0^0(980)f_{0}(500)}\cdot\frac{\Pi_{a_0^0f_0}(q^2)}{D_{a_0^0}(q^2)D_{f_0}(q^2)-\Pi_{a_0^0f_0}^2(q^2)}
\cdot g_{f_0(980)\pi^+\pi^-},
\end{align}
here, $q^2=(p_{\pi^+}+p_{\pi^-})^2$, $a_0^0$ and $f_0$ is respectively the shorthand of $a_0^0(980)$ and $f_0(980)$. ${\mathcal M}_{a_0(1450)a_0^0(980)f_{0}(500)}$ is the invariant amplitude for the decay $a_0(1450)\rightarrow a_0^0(980)f_{0}(500)$,
\begin{align}\label{eq:decayamplitudeoaasig}
&{\mathcal B}(a_0(1450)\rightarrow a_0^0(980)f_{0}(500))=
\left|{\mathcal M}_{a_0(1450)a_0^0(980)f_{0}(500)}\right|^2\cdot\frac{f(m_{a_0(1450)},m_{a_0^0(980)},m_{f_0(500)})}{16\pi\Gamma_{a_0(1450)}m_{a_0(1450)}^3},
\end{align}
hereinafter,$m_{r}$ denote the mass of resonance r $\left[r=f_0(980)\right.$, $ a_0^0(980)$, $f_0(500)$, $a_0(1450)$, $\eta$, $\pi^0$, $\pi^+$, $K^0$, $\left. K^+ \right]$ and $\Gamma_{r}$ with  $r=a_0^0(980),f_0(980),a_0(1450)$ denote the width of the resonance. The function $f(x,y,z)$ is defined as
\begin{align}\label{eq:funcfxyz}
f(x,y,z)=\sqrt{{x^4} +{y^4}+{z^4}-2{x^2}{y^2}-2{x^2}{z^2}- 2{y^2}{z^2} },
\end{align}
$g_{f_0(980)\pi^+\pi^-}$ is the coupling constant of $f_0(980)$ with $\pi^+\pi^-$ and can be extracted from the branching ratio of the $f_0(980)\rightarrow\pi^+\pi^-$ decay
\begin{align}\label{eq:branrafnepipi}
&{\mathcal B}(f_0(980)\rightarrow \pi^+\pi^-)=
\left|g_{f_0(980)\pi^+\pi^-}\right|^2\cdot\frac{f(m_{f_0(980)},m_{\pi^+},m_{\pi^+})}{16\pi\Gamma_{f_0(980)}m_{f_0(980)}^3}.
\end{align}
The $a_0^0(980)$-$f_0(980)$ mixing amplitude $\Pi_{a_0^0f_0}(q^2)$ has the following form~\cite{Achasov:2017ncx,Achasov:2018grq}
\begin{align}\label{eq:decayamplioffamixng}
\Pi_{a_0^0f_0}(q^2)=&\frac{g_{a_0^0(980)K^{+}K^{-}}g_{f_0(980)K^{+}K^{-}}}{16\pi}\bigg[i\left(R_{K^{+}K^{-}}(q^2)-R_{K^{0}{\bar{K}}^{0}}(q^2)\right)\bigg.\nonumber\\
&\bigg.-\frac{R_{K^{+}K^{-}}(q^2)}{\pi} \ln{\frac{1+R_{K^{+}K^{-}}(q^2)}{1-R_{K^{+}K^{-}}(q^2)}}+\frac{R_{K^{0}{\bar{K}}^{0}}(q^2)}{\pi} \ln{\frac{1+R_{K^{0}{\bar{K}}^{0}}(q^2)}{1-R_{K^{0}{\bar{K}}^{0}}(q^2)}}\bigg],
\end{align}
where $g_{a_0^0(980)K^{+}K^{-}}$ and $g_{f_0(980)K^{+}K^{-}}$ is the coupling constant of $K^+K^-$ with $a_0^0(980)$ and $f_0(980)$, respectively. For $q^2\geq 4 m_a^2$ $[a=K^+,K^0]$, $R_{aa}(q^2)=\sqrt{1-{4 m_a^2}/{q^2}}$, if  $q^2\le 4 m_a^2$, then $R_{aa}(q^2)$ should be replaced by $i\sqrt{{4 m_a^2}/{q^2}-1}$.
In Eq.(\ref{eq:decayamplitudeofazzfa}), $D_{r}(q^2)$ is the inverse propagator of the unmixed resonance $r$,
\begin{align}\label{eq:decaywidthoffamixng3}
D_{r}(q^2)=q^2-m_r^2-\sum\limits_{ab}\left[\text{Re}\Pi_r^{ab}(m_r^2)-\Pi_r^{ab}(q^2)\right].
\end{align}
For $r=a_0^0(980)$, $ab=\left(\eta\pi^0,K^+k^-,K^0{\bar{K}}^0\right)$; for $r=f_0(980)$, $ab=\left(\pi^+\pi^-,\pi^0\pi^0,K^+k^-,K^0{\bar{K}}^0\right)$. $\Pi_r^{ab}$ denote the diagonal matrix of the polarization operator of the resonance $r$ corresponding to the one loop contribution from the two-particle intermediate $ab$ states, it is a piecewise function, its expressions in the different $q^2$ regions are displayed in Eqs.(18-20) of Ref.~\cite{Cheng:2018smm}. Making use of Eqs.(\ref{eq:decayamplitudeofazzfa}), (\ref{eq:decayamplitudeoaasig}) and (\ref{eq:branrafnepipi}), it is then straightforward to obtain
\begin{align}\label{eq:azzdewidoffazzmixng1}
&\frac{d\Gamma(a_0(1450)\rightarrow a_0^0(980)f_{0}(500)\rightarrow f_0(980)f_{0}(500)\rightarrow\pi^+ \pi^- f_{0}(500))}{dq^2}={\mathcal B}(f_0(980)\rightarrow\pi^+ \pi^-)\nonumber\\
&~~~~~~~~~~~~~~~~~~~~~~~~~\cdot{\mathcal B}(a_0(1450)\rightarrow a_0^0(980)f_{0}(500))\cdot \left|\frac{\Pi_{a_0^0f_0}(q^2)}{D_{a_0^0}(q^2)D_{f_0}(q^2)-\Pi_{a_0^0f_0}^2(q^2)}\right|^2 \cdot\varphi_{S},
\end{align}
where $\varphi_{S}$ is the relevant phase-space factor
\begin{align}\label{eq:azzdewidoffazzmixng2}
\varphi_{S}=\frac{\Gamma_{a_0(1450)}\Gamma_{f_0(980)} m_{f_0(980)}^3}{\pi q^2}\cdot\frac{f(m_{a_0(1450)},m_{f_0(500)},\sqrt{q^2})}{f(m_{a_0(1450)},m_{f_0(500)},m_{a_0^0(980)})} \cdot \frac{f(\sqrt{q^2},m_{\pi^+},m_{\pi^+})}{f(m_{f_0(980)},m_{\pi^+},m_{\pi^+})},
\end{align}
By multiplying both sides of Eq.(\ref{eq:azzdewidoffazzmixng1}) by ${\mathcal B}(J/\psi\rightarrow \gamma \eta_c)$ and ${\mathcal B}(\eta_c \rightarrow a_0(1450)\pi^0)$, one can obtain
\begin{align}\label{eq:azzdewidoffazzmixng3}
&{\mathcal B}(J/\psi\rightarrow \gamma \eta_c\rightarrow \gamma a_0(1450)\pi^0)\cdot\frac{d\Gamma(a_0(1450)\rightarrow a_0^0(980)f_{0}(500)\rightarrow\pi^+ \pi^- f_{0}(500))}{dq^2}\nonumber\\
&={\mathcal B}(J/\psi\rightarrow \gamma \eta_c)\cdot {\mathcal B}(\eta_c \rightarrow a_0(1450)\pi^0)\cdot{\mathcal B}(a_0(1450)\rightarrow a_0^0(980)f_{0}(500))\nonumber\\
&\cdot {\mathcal B}(f_0(980)\rightarrow\pi^+ \pi^-)\cdot \left|\frac{\Pi_{a_0^0f_0}(q^2)}{D_{a_0^0}(q^2)D_{f_0}(q^2)-\Pi_{a_0^0f_0}^2(q^2)}\right|^2 \cdot\varphi_{S},
\end{align}

\begin{figure}[t]
\centering
\includegraphics[width=0.42\textwidth]{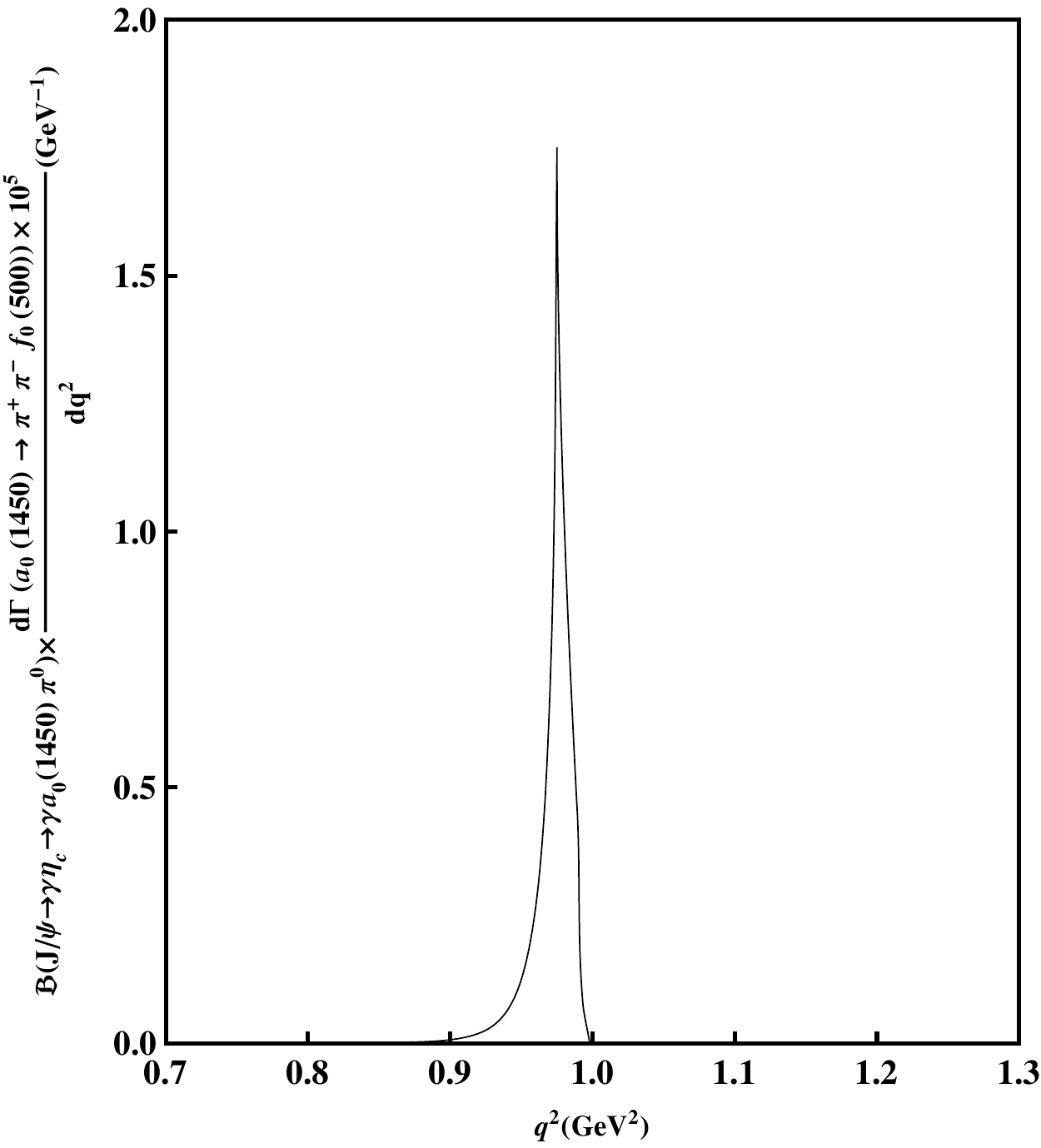}
\caption{\small the distribution of the $\pi^+\pi^-$ mass square spectrum  $(q^2=(p_{\pi^+}+p_{\pi^-})^2)$ for the decay $J/\psi \rightarrow\gamma\eta_c \rightarrow \gamma \pi^0 a_0^0(1450)\rightarrow \gamma \pi^0 a_0^0(980)f_0(500) \rightarrow \gamma \pi^0 \pi^+\pi^- f_0(500)$.}
\label{maffsquaredistribution}
\end{figure}

With the value of the parameters which are listed in Table~\ref{inputmassparameter} and substituting Eq.(\ref{eq:pdgetctokkpiz1}), Eq.(\ref{eq:brchainetacasi}) and Eq.(\ref{eq:branchingratioofa9802}) into Eq.(\ref{eq:azzdewidoffazzmixng3}), we can obtain the distribution curve of the $\pi^+\pi^-$ mass square spectrum for the decay $J/\psi \rightarrow\gamma\eta_c \rightarrow \gamma \pi^0 a_0^0(1450)\rightarrow \gamma \pi^0 a_0^0(980)f_0(500) \rightarrow \gamma \pi^0 \pi^+\pi^- f_0(500)$, which is presented in Fig.~\ref{maffsquaredistribution}. Here, we note that the narrow peak from the $a_0^0(980)$-$f_{0}(980)$ mixing can be clearly observed in this figure.
\begin{table}[t]
\begin{center}
\caption{\label{inputmassparameter} \small Properties of the resonances.}
\vspace{0.2cm}
\doublerulesep 0.8pt \tabcolsep 0.18in
\begin{tabular}{cc}
\hline
$m_{\pi^+}=  139.6 \text{MeV}$ ~\cite{Tanabashi:2018oca}                         &$m_{\pi^0}=  135 \text{MeV}$ ~\cite{Tanabashi:2018oca}  \\
\hline
$m_{K^+}=  493.7 \text{MeV}$  ~\cite{Tanabashi:2018oca}                        &$m_{K^0}=  497.6 \text{MeV}$ ~\cite{Tanabashi:2018oca}     \\
\hline
$m_{\eta}=  547.9 \text{MeV}$~\cite{Tanabashi:2018oca}                &$m_{\eta^{\prime}}=  (957.8\pm 0.1) \text{MeV}$~\cite{Tanabashi:2018oca}   \\
\hline
$m_{f_0(980)}=  (0.99\pm 0.02) \text{GeV}$ ~\cite{Tanabashi:2018oca}                         &${\Gamma}_{f_0(980)}=  0.074 \text{GeV}$~\cite{Ablikim:2017auj}  \\
\hline
$m_{a_0^0(980)}=  (0.98\pm 0.02) \text{GeV}$  ~\cite{Tanabashi:2018oca}         &${\Gamma}_{a_0^0(980)}=  (0.092\pm 0.008) \text{GeV}$~\cite{Tanabashi:2018oca}     \\
\hline
$m_{a_0(1450)}=  (1.474\pm 0.019) \text{GeV}$~\cite{Tanabashi:2018oca}               &${\Gamma}_{a_0(1450)}=  (0.265\pm 0.013) \text{GeV} $~\cite{Tanabashi:2018oca} \\
\hline
$m_{f_0(500)}=  (0.475\pm 0.075) \text{GeV}$~\cite{Tanabashi:2018oca}               &$g_{a_0^0(980)\eta\pi^0}=2.43 \text{GeV}$  ~\cite{Cheng:2005nb,Cheng:2013fba}          \\
\hline
$g_{a_0^0(980)K^{+}K^{-}}=(2.76\pm 0.46) \text{GeV}$~\cite{Kornicer:2016axs,Ablikim:2018ffp}                 &$g_{a_0^0(980)K^{0}{\bar{K}}^{0}}=(2.76\pm 0.46) \text{GeV}$~\cite{Kornicer:2016axs,Ablikim:2018ffp}\\
\hline
$g_{f_0(980)\pi^{+}\pi^{-}}=1.39 \text{GeV}$~\cite{Cheng:2005nb,Cheng:2013fba}          &$g_{f_0(980)\pi^{0}\pi^{0}}=0.98 \text{GeV}$~\cite{Cheng:2005nb,Cheng:2013fba}           \\
\hline
$g_{f_0(980)K^{+}K^{-}}=3.17 \text{GeV}$~\cite{Achasov:2017ncx}                 &$g_{f_0(980)K^{0}{\bar{K}}^{0}}=3.17 \text{GeV}$~\cite{Achasov:2017ncx}      \\
\hline
\end{tabular}
\end{center}
\end{table}
The physical range of $q^2$ for $a_0^0(1450)\rightarrow\pi^+\pi^- f_0(500)$ is $4m_{\pi^+}^{2}\leq q^2\leq (m_{a_0(1450)}-m_{f_0(500)})^2$. By integrating over the variable $q^2$, we finally obtain the following value of the branching ratio
\begin{align}\label{eq:branchingratioofthetotal}
{{\mathcal B}( J/\psi \rightarrow\gamma\eta_c \rightarrow \gamma \pi^0 a_0^0(1450)\rightarrow \gamma \pi^0 a_0^0(980)f_0(500) \rightarrow \gamma \pi^0 \pi^+\pi^- f_0(500))=\left(1.04^{+0.68}_{-0.61}\right)\times 10^{-6},}
\end{align}
here, we combine the uncertainties of the branching ratios involved in the calculation, the decay width of $a_0(1450)$ and the mass of $f_0(500)$ to determine the final error of the aboved branching ratio. Adding Eq.(\ref{eq:branchingratioofa9801}) and Eq.(\ref{eq:branchingratioofthetotal}) together, we then arrive at
\begin{align}\label{eq:branchingratioofthetotpipi}
{{\mathcal B}( J/\psi \rightarrow\gamma\eta_c \rightarrow \gamma \pi^0 a_0^0(1450)\rightarrow \gamma \pi^0 a_0^0(980)f_0(500) \rightarrow \gamma \pi^0 \pi^+\pi^- \pi^+\pi^-)=\left(0.70^{+0.46}_{-0.41}\right)\times 10^{-6},}
\end{align}
\section{the non-resonant decay $a_0^0(1450)\rightarrow  f_0(980) \pi^+\pi^-$ }
\label{sec:nonresonantdecay}
In experiment, the resonant $f_0(500)$ is reconstructed by the $f_0(500)\rightarrow \pi^+\pi^-$ decay. If we apply the selection criteria which restrict the invariant mass of $\pi^+\pi^-$ to the $f_0(500)$ mass window, the background channel $a_0^0(1450)\rightarrow  f_0(980) \pi^+\pi^-(\text{non-resonant})$ can not be removed because of the large width of the resonant $f_0(500)$. Fortunately, however, the the non-resonant decay  $a_0^0(1450)\rightarrow  f_0(980) \pi^+\pi^-$ violates isospin invariant or C symmetry, the violation of isospin invariant is caused by the $a_0^0(980)$-$f_{0}(980)$ mixing.
 
In the isospin limit, the wave function of the two pions system can be written as~\cite{Buchoff:2008hh,Li:2009jd}
\begin{align}
&(\pi\pi)^{I_{3}=0}_{I=0}=\frac{\sqrt{3}}{3}\left| \pi^+ \right\rangle \left| \pi^- \right\rangle +\frac{\sqrt{3}}{3}\left| \pi^- \right\rangle \left| \pi^+\right\rangle -\frac{\sqrt{3}}{3}\left| \pi^0 \right\rangle \left| \pi^0\right\rangle\label{eq:pipiisospin1},\\
&(\pi\pi)^{I_{3}=0}_{I=1}=\frac{1}{\sqrt{2}}\left| \pi^+ \right\rangle \left| \pi^- \right\rangle -\frac{1}{\sqrt{2}}\left| \pi^- \right\rangle \left| \pi^+\right\rangle\label{eq:pipiisospin2},\\
&(\pi\pi)^{I_{3}=0}_{I=2}=\frac{\sqrt{6}}{6}\left| \pi^+ \right\rangle \left| \pi^- \right\rangle +\frac{\sqrt{6}}{6}\left| \pi^- \right\rangle \left| \pi^+\right\rangle +\frac{\sqrt{6}}{3}\left| \pi^0 \right\rangle \left| \pi^0\right\rangle\label{eq:pipiisospin3},
\end{align}
In the above equations, we can see that the C parity of $(\pi\pi)^{I_{3}=0}_{I=0}$, $(\pi\pi)^{I_{3}=0}_{I=1}$, and $(\pi\pi)^{I_{3}=0}_{I=2}$ are $+1$, $-1$ and $+1$, respectively. As for the non-resonant decay  $a_0^0(1450)\rightarrow  f_0(980) \pi^+\pi^-$, if isospin is conserved, two pions system have $I=1,I_3=0$, so the the C parity of two pions system is $-1$, as a consequence, $C(f_0(980) (\pi\pi)^{I_{3}=0}_{I=1})=-(f_0(980) (\pi\pi)^{I_{3}=0}_{I=1})$, while it is $C=+1$ for $a_0^0(1450)$, therefore this decay violate C. if the non-resonant decay  $a_0^0(1450)\rightarrow  f_0(980) \pi^+\pi^-$ violate isospin, two pions system have $I=2,I_3=0$ or
$I=0,I_3=0$, then the the C parity of two pions system is $+1$, we can easily achieve that this decay conserve C. In a word, the non-resonant decay  $a_0^0(1450)\rightarrow  f_0(980) \pi^+\pi^-$ violates C or I. Because C violation is only known to occur in weak interaction, the contribution from C violation is much smaller than that from the isospin violation, which can occur in electromagnetic interaction, so the contribution from C violation can be neglected. The non-resonant decay  $a_0^0(1450)\rightarrow  f_0(980) \pi^+\pi^-$ is determined mainly by the contribution of the isospin symmetry breaking process which is cased by the $a_0^0(980)$-$f_{0}(980)$ mixing, so the non-resonant decay $a_0^0(1450)\rightarrow  f_0(980) \pi^+\pi^-$ and the decay $a_0^0(1450)\rightarrow  f_0(980) f_0(500)$ can be combined to measure the mixing intensity $\xi_{af}$ in experiment.
\section{{Prospects for the measurement at BESIII} }
As for the decay $J/\psi \rightarrow\gamma\eta_c \rightarrow \gamma \pi^0 a_0^0(1450)\rightarrow \gamma \pi^0 a_0^0(980)f_0(500)\rightarrow \gamma \pi^0 f_0(980) f_0(500) \rightarrow \gamma \pi^0 \pi^+\pi^- \pi^+\pi^-$, there are four intermediate states, i.e., $\eta_c$, $a_0^0(1450)$, $f_0(500)$ and $f_0(980)$. Because of the narrow peak near the $K\bar{K}$ thresholds in the $\pi^+\pi^-$ invariant mass spectrum, the event selection criteria for the $f_0 (980)$ candidates has high efficiency. As discussed in section~\ref{sec:nonresonantdecay}, the selection criteria which constraint the invariant mass of $\pi^+\pi^-$ to the $f_0(500)$ mass window also has high efficiency when both the non-resonant decay $a_0^0(1450)\rightarrow  f_0(980) \pi^+\pi^-$ and the decay $a_0^0(1450)\rightarrow  f_0(980) f_0(500)$ are combined. In addition, the $\pi^0$ final state is reconstructed through the decay $\pi^0\rightarrow\gamma\gamma$, which branching ratio is $(98.82\pm0.03)\%$~\cite{Tanabashi:2018oca}, so the final states of the decay $J/\psi \rightarrow\gamma\eta_c \rightarrow \gamma \pi^0 a_0^0(1450)\rightarrow \gamma \pi^0 a_0^0(980)f_0(500)\rightarrow \gamma \pi^0 f_0(980) f_0(500) \rightarrow \gamma \pi^0 \pi^+\pi^- \pi^+\pi^-$ contain three photons and four charged track. After considering all these above, we assume that the efficiency for $J/\psi \rightarrow\gamma\eta_c \rightarrow \gamma \pi^0 a_0^0(1450)\rightarrow \gamma \pi^0 a_0^0(980)f_0(500)\rightarrow \gamma \pi^0 f_0(980) f_0(500) \rightarrow \gamma \pi^0 \pi^+\pi^- \pi^+\pi^-$ is $3\%$ after the final selection~\cite{Ablikim:2012ur,Ablikim:2014pfc,Ablikim:2016frj,Asner:2008nq}, so the branching ratio-times-efficiency factor of this decay can reach about $3.0\times 10^{-8}$. The BESIII experiment will produce $10\times 10^9$ $J/\psi$ events~\cite{Asner:2008nq,Li:2016tlt,Bigi:2017eni,Fang:2017qgz}, so about 300 events should be observed in the corresponding signal region. Therefore, the isospin breaking decay $J/\psi \rightarrow\gamma\eta_c \rightarrow \gamma \pi^0 a_0^0(1450)\rightarrow \gamma \pi^0 a_0^0(980)f_0(500)\rightarrow \gamma \pi^0 f_0(980) f_0(500) \rightarrow \gamma \pi^0 \pi^+\pi^- \pi^+\pi^-$ may be used to study the $a_0^0(980)$-$f_{0}(980)$ mixing and determine the value of $\xi_{af}$ exactly.
\section{Conclusions}
\label{sec:Conclusions}
In summary, using the $f_0(980)$-$a_0^0(980)$ mixing intensity $\xi_{af}$ measured recently at BESIII~\cite{Ablikim:2018pik}, we investigate the $f_0(980)$-$a_0^0(980)$ mixing through the isospin violation decay $J/\psi \rightarrow\gamma\eta_c \rightarrow \gamma \pi^0 a_0^0(1450)\rightarrow \gamma \pi^0 a_0^0(980)f_0(500)\rightarrow \gamma \pi^0 f_0(980) f_0(500) \rightarrow \gamma \pi^0 \pi^+\pi^- \pi^+\pi^-$. We find that the branching ratio for the decay can reach up to the order of $10^{-6}$, which might be hopefully measurable with $10^{10}$ $J/\psi$ events collected at BESIII. We also observe the narrow peak from the $f_0(980)$-$a_0^0(980)$ mixing in the $\pi^+\pi^-$ mass spectrum. The  related decay $a_0^0(1450)\rightarrow  f_0(980) \pi^+\pi^-\text{(non-resonant)}$ which is dominated by the contribution of the isospin symmetry breaking process can be combined with the decay $a_0^0(1450)\rightarrow a_0^0(980)f_0(500)\rightarrow f_0(980) f_0(500) \rightarrow \pi^+\pi^- \pi^+\pi^-$ to study the $f_0(980)$-$a_0^0(980)$ mixing in experiment. These decays could be complementary to the decay $\chi_{c1}\rightarrow f_{0}(980)\pi^{0}\to\pi^{+}\pi^{-}\pi^{0}$ which had been observed at BESIII~\cite{Ablikim:2018pik}, the observations of them will make the measurement of the mixing intensity $\xi_{af}$ more precisely, and then  enhancing the understanding the nature of the light scalar mesons.
\section*{Acknowledgements}
The work was supported by the National Natural Science Foundation of China (Contract No. 11675137) and the Key Scientific Research Projects of Colleges and Universities in Henan Province (Contract No. 18A140029).

\begin{appendix}

\end{appendix}


\begin{thebibliography}{100}

\bibitem{Cheng:2005nb}
  H.~Y.~Cheng, C.~K.~Chua and K.~C.~Yang,
  Phys.\ Rev.\ D {\bf 73}, 014017 (2006)
  doi:10.1103/PhysRevD.73.014017
  [hep-ph/0508104].

\bibitem{Weinstein:1983gd}
  J.~D.~Weinstein and N.~Isgur,
  Phys.\ Rev.\  D {\bf 27} (1983) 588.

\bibitem{Weinstein:1982gc}
  J.~D.~Weinstein and N.~Isgur,
  Phys.\ Rev.\ Lett.\  {\bf 48} (1982) 659.

\bibitem{Weinstein:1990gu}
  J.~D.~Weinstein and N.~Isgur,
  Phys.\ Rev.\  D {\bf 41} (1990) 2236.

\bibitem{Jaffe} R.\ L. Jaffe, Phys.\ Rev.\ D15 (1977) 267; ibid. (1977) 281.

\bibitem{Kim:2017yur}
  K.~S.~Kim and H.~Kim,
  Eur.\ Phys.\ J.\ C {\bf 77}, no. 7, 435 (2017)
  doi:10.1140/epjc/s10052-017-5020-5
  [arXiv:1703.01390 [hep-ph]].

\bibitem{Abdel-Rehim:2014zwa}
  J.~Berlin, A.~Abdel-Rehim, C.~Alexandrou, M.~Dalla Brida, M.~Gravina and M.~Wagner,
  PoS LATTICE {\bf 2014}, 104 (2014)
  [arXiv:1410.8757 [hep-lat]].

\bibitem{Amsler:1995tu}
  C.~Amsler and F.~E.~Close,
  Phys.\ Lett.\  B {\bf 353} (1995) 385
  [arXiv:hep-ph/9505219].

\bibitem{Amsler:1995td}
  C.~Amsler and F.~E.~Close,
  Phys.\ Rev.\  D {\bf 53} (1996) 295
  [arXiv:hep-ph/9507326].

\bibitem{Amsler:2002ey}
  C.~Amsler,
  Phys.\ Lett.\  B {\bf 541} (2002) 22
  [arXiv:hep-ph/0206104].

\bibitem{Gorishnii:1983zi}
  S.~G.~Gorishnii, A.~L.~Kataev and S.~A.~Larin,
  Phys.\ Lett.\  {\bf 135B}, 457 (1984).
  doi:10.1016/0370-2693(84)90315-0;
  A.~L.~Kataev,
  Phys.\ Atom.\ Nucl.\  {\bf 68}, 567 (2005)
  [Yad.\ Fiz.\  {\bf 68}, 597 (2005)]
  doi:10.1134/1.1903086
  [hep-ph/0406305].

\bibitem{Achasov:1979xc}
  N.~N.~Achasov, S.~A.~Devyanin and G.~N.~Shestakov,
  Phys.\ Lett.\  {\bf 88B}, 367 (1979).
  doi:10.1016/0370-2693(79)90488-X

\bibitem{Achasov:1981de}
  N.~N.~Achasov, S.~A.~Devyanin and G.~N.~Shestakov,
  Yad.\ Fiz.\  {\bf 33}, 1337 (1981)
  [Sov.\ J.\ Nucl.\ Phys.\  {\bf 33}, 715 (1981)].

\bibitem{Achasov:1996qn}
  N.~N.~Achasov and G.~N.~Shestakov,
  Phys.\ Rev.\ D {\bf 56}, 212 (1997)
  doi:10.1103/PhysRevD.56.212
  [hep-ph/9610409].

\bibitem{Krehl:1996rk}
  O.~Krehl, R.~Rapp and J.~Speth,
  Phys.\ Lett.\ B {\bf 390}, 23 (1997)
  doi:10.1016/S0370-2693(96)01425-6
  [nucl-th/9609013].

\bibitem{Kerbikov:2000pu}
  B.~Kerbikov and F.~Tabakin,
  Phys.\ Rev.\ C {\bf 62}, 064601 (2000)
  doi:10.1103/PhysRevC.62.064601
  [nucl-th/0006017].

\bibitem{Close:2000ah}
  F.~E.~Close and A.~Kirk,
  Phys.\ Lett.\ B {\bf 489}, 24 (2000)
  doi:10.1016/S0370-2693(00)00951-5
  [hep-ph/0008066].

\bibitem{Kudryavtsev:2001ee}
  A.~E.~Kudryavtsev and V.~E.~Tarasov,
  JETP Lett.\  {\bf 72}, 410 (2000)
  [Pisma Zh.\ Eksp.\ Teor.\ Fiz.\  {\bf 72}, 589 (2000)]
  doi:10.1134/1.1335118
  [nucl-th/0102053].

\bibitem{Grishina:2001zj}
  V.~Y.~Grishina, L.~A.~Kondratyuk, M.~Buescher, W.~Cassing and H.~Stroher,
  Phys.\ Lett.\ B {\bf 521}, 217 (2001)
  doi:10.1016/S0370-2693(01)01210-2
  [nucl-th/0103081].

\bibitem{Close:2001ay}
  F.~E.~Close and A.~Kirk,
  Phys.\ Lett.\ B {\bf 515}, 13 (2001)
  doi:10.1016/S0370-2693(01)00799-7
  [hep-ph/0106108].

\bibitem{Kudryavtsev:2002uu}
  A.~E.~Kudryavtsev, V.~E.~Tarasov, J.~Haidenbauer, C.~Hanhart and J.~Speth,
  Phys.\ Rev.\ C {\bf 66}, 015207 (2002)
  doi:10.1103/PhysRevC.66.015207
  [nucl-th/0203034].

\bibitem{Kondratyuk:2002yf}
  L.~A.~Kondratyuk, E.~L.~Bratkovskaya, V.~Y.~Grishina, M.~Buescher, W.~Cassing and H.~Stroher,
  Phys.\ Atom.\ Nucl.\  {\bf 66}, 152 (2003)
  [Yad.\ Fiz.\  {\bf 66}, 155 (2003)]
  doi:10.1134/1.1540670
  [nucl-th/0207033].

\bibitem{Achasov:2002hg}
  N.~N.~Achasov and A.~V.~Kiselev,
  Phys.\ Lett.\ B {\bf 534}, 83 (2002)
  doi:10.1016/S0370-2693(02)01696-9
  [hep-ph/0203042].

\bibitem{Achasov:2003se}
  N.~N.~Achasov and G.~N.~Shestakov,
  Phys.\ Rev.\ Lett.\  {\bf 92}, 182001 (2004)
  doi:10.1103/PhysRevLett.92.182001
  [hep-ph/0312214].

\bibitem{Grishina:2004rd}
  V.~Y.~Grishina, L.~A.~Kondratyuk, M.~Buescher and W.~Cassing,
  Eur.\ Phys.\ J.\ A {\bf 21}, 507 (2004)
  doi:10.1140/epja/i2004-10004-2
  [nucl-th/0402093].

\bibitem{Achasov:2004ur}
  N.~N.~Achasov and G.~N.~Shestakov,
  Phys.\ Rev.\ D {\bf 70}, 074015 (2004)
  doi:10.1103/PhysRevD.70.074015
  [hep-ph/0405129].

\bibitem{Wu:2007jh}
  J.~J.~Wu, Q.~Zhao and B.~S.~Zou,
  Phys.\ Rev.\ D {\bf 75}, 114012 (2007)
  doi:10.1103/PhysRevD.75.114012
  [arXiv:0704.3652 [hep-ph]].

\bibitem{Wu:2008hx}
  J.~J.~Wu and B.~S.~Zou,
  Phys.\ Rev.\ D {\bf 78}, 074017 (2008)
  doi:10.1103/PhysRevD.78.074017
  [arXiv:0808.2683 [hep-ph]].

\bibitem{Hanhart:2007bd}
  C.~Hanhart, B.~Kubis and J.~R.~Pelaez,
  Phys.\ Rev.\ D {\bf 76}, 074028 (2007)
  doi:10.1103/PhysRevD.76.074028
  [arXiv:0707.0262 [hep-ph]].

\bibitem{Ablikim:2010aa}
  M.~Ablikim {\it et al.} [BESIII Collaboration],
  Phys.\ Rev.\ D {\bf 83}, 032003 (2011)
  doi:10.1103/PhysRevD.83.032003
  [arXiv:1012.5131 [hep-ex]].

\bibitem{Wu:2011yx}
  J.~J.~Wu, X.~H.~Liu, Q.~Zhao and B.~S.~Zou,
  Phys.\ Rev.\ Lett.\  {\bf 108}, 081803 (2012)
  doi:10.1103/PhysRevLett.108.081803
  [arXiv:1108.3772 [hep-ph]].

\bibitem{Aceti:2012dj}
  F.~Aceti, W.~H.~Liang, E.~Oset, J.~J.~Wu and B.~S.~Zou,
  Phys.\ Rev.\ D {\bf 86}, 114007 (2012)
  doi:10.1103/PhysRevD.86.114007
  [arXiv:1209.6507 [hep-ph]].

\bibitem{Roca:2012cv}
  L.~Roca,
  Phys.\ Rev.\ D {\bf 88}, 014045 (2013)
  doi:10.1103/PhysRevD.88.014045
  [arXiv:1210.4742 [hep-ph], arXiv:1210.4742 [hep-ph]].

\bibitem{Tarasov:2013yma}
  V.~E.~Tarasov, W.~J.~Briscoe, W.~Gradl, A.~E.~Kudryavtsev and I.~I.~Strakovsky,
  Phys.\ Rev.\ C {\bf 88}, 035207 (2013)
  doi:10.1103/PhysRevC.88.035207
  [arXiv:1306.6618 [hep-ph]].

\bibitem{Sekihara:2014qxa}
  T.~Sekihara and S.~Kumano,
  Phys.\ Rev.\ D {\bf 92}, no. 3, 034010 (2015)
  doi:10.1103/PhysRevD.92.034010
  [arXiv:1409.2213 [hep-ph]].

\bibitem{Aceti:2015zva}
  F.~Aceti, J.~M.~Dias and E.~Oset,
  Eur.\ Phys.\ J.\ A {\bf 51}, no. 4, 48 (2015)
  doi:10.1140/epja/i2015-15048-5
  [arXiv:1501.06505 [hep-ph]].

\bibitem{Achasov:2015uua}
  N.~N.~Achasov, A.~A.~Kozhevnikov and G.~N.~Shestakov,
  Phys.\ Rev.\ D {\bf 92}, no. 3, 036003 (2015)
  doi:10.1103/PhysRevD.92.036003
  [arXiv:1504.02844 [hep-ph]].

\bibitem{Wang:2016wpc}
  W.~Wang,
  Phys.\ Lett.\ B {\bf 759}, 501 (2016)
  doi:10.1016/j.physletb.2016.06.007
  [arXiv:1602.05288 [hep-ph]].

\bibitem{Achasov:2016wll}
  N.~N.~Achasov, A.~A.~Kozhevnikov and G.~N.~Shestakov,
  Phys.\ Rev.\ D {\bf 93}, no. 11, 114027 (2016)
  doi:10.1103/PhysRevD.93.114027
  [arXiv:1604.00177 [hep-ph]].

\bibitem{Achasov:2017edm}
  N.~N.~Achasov and G.~N.~Shestakov,
  Phys.\ Rev.\ D {\bf 96}, no. 3, 036013 (2017)
  doi:10.1103/PhysRevD.96.036013
  [arXiv:1704.01763 [hep-ph]].

\bibitem{Achasov:2017zhu}
  N.~N.~Achasov and G.~N.~Shestakov,
  Phys.\ Rev.\ D {\bf 96}, no. 1, 016027 (2017)
  doi:10.1103/PhysRevD.96.016027
  [arXiv:1705.08689 [hep-ph]].

\bibitem{Sakai:2017iqs}
  S.~Sakai, E.~Oset and W.~H.~Liang,
  Phys.\ Rev.\ D {\bf 96}, no. 7, 074025 (2017)
  doi:10.1103/PhysRevD.96.074025
  [arXiv:1707.02236 [hep-ph]].

\bibitem{Bayar:2017pzq}
  M.~Bayar and V.~R.~Debastiani,
  Phys.\ Lett.\ B {\bf 775}, 94 (2017)
  doi:10.1016/j.physletb.2017.10.061
  [arXiv:1708.02764 [hep-ph]].

\bibitem{Achasov:2017ncx}
  N.~N.~Achasov and G.~N.~Shestakov,
  Phys.\ Rev.\ D {\bf 96}, no. 9, 091501 (2017)
  doi:10.1103/PhysRevD.96.091501
  [arXiv:1709.02068 [hep-ph]].

\bibitem{Liang:2017ijf}
  W.~H.~Liang, S.~Sakai, J.~J.~Xie and E.~Oset,
  Chin.\ Phys.\ C {\bf 42}, no. 4, 044101 (2018)
  doi:10.1088/1674-1137/42/4/044101
  [arXiv:1711.04603 [hep-ph]].

\bibitem{Achasov:2019ywj}
 N.~Achasov and G.~Shestakov,
 Usp. Fiz. Nauk \textbf{189}, no.1, 3-32 (2019)
 doi:10.3367/UFNr.2018.01.038281

\bibitem{Achasov:2019zrc}
 N.~Achasov and G.~Shestakov,
 EPJ Web Conf. \textbf{212}, 03002 (2019)
 doi:10.1051/epjconf/201921203002

\bibitem{Liang:2019jtr}
W.~H.~Liang, H.~X.~Chen, E.~Oset and E.~Wang,
Eur. Phys. J. C \textbf{79}, no.5, 411 (2019)
doi:10.1140/epjc/s10052-019-6928-8
[arXiv:1903.01252 [hep-ph]].

\bibitem{Molina:2019udw}
 R.~Molina, J.~J.~Xie, W.~H.~Liang, L.~S.~Geng and E.~Oset,
 Phys. Lett. B \textbf{803}, 135279 (2020)
 doi:10.1016/j.physletb.2020.135279
 [arXiv:1908.11557 [hep-ph]].

\bibitem{Achasov:2020fee}
 N.~Achasov,
 [arXiv:2002.01354 [hep-ph]].

\bibitem{Achasov:2020qfx}
  N.~Achasov, A.~Kiselev and G.~Shestakov,
  [arXiv:2005.06455 [hep-ph]].

\bibitem{Ablikim:2018pik}
 M.~Ablikim \textit{et al.} [BESIII],
 Phys. Rev. Lett. \textbf{121}, no.2, 022001 (2018)
 doi:10.1103/PhysRevLett.121.022001
 [arXiv:1802.00583 [hep-ex]].

\bibitem{Aliev:2018bln}
  T.~M.~Aliev and S.~Bilmis,
  Eur.\ Phys.\ J.\ A {\bf 54}, no. 9, 147 (2018)
  doi:10.1140/epja/i2018-12584-4
  [arXiv:1808.08843 [hep-ph]].

\bibitem{Mathur:2006bs}
 N.~Mathur, A.~Alexandru, Y.~Chen, S.~Dong, T.~Draper, I.~Horvath, F.~Lee, K.~Liu, S.~Tamhankar and J.~Zhang,
 Phys. Rev. D \textbf{76} (2007), 114505
 doi:10.1103/PhysRevD.76.114505
 [arXiv:hep-ph/0607110 [hep-ph]].

\bibitem{lee}
 W.~J.~Lee and D.~Weingarten,
 Phys. Rev. D \textbf{61}, 014015 (2000)
 doi:10.1103/PhysRevD.61.014015
 [arXiv:hep-lat/9910008 [hep-lat]].
 M.~Gockeler, R.~Horsley, H.~Perlt, P.~E.~Rakow, G.~Schierholz, A.~Schiller and P.~Stephenson,
 Phys. Rev. D \textbf{57}, 5562-5580 (1998)
 doi:10.1103/PhysRevD.57.5562
 [arXiv:hep-lat/9707021 [hep-lat]].
 S.~Kim and S.~Ohta,
 Nucl. Phys. B Proc. Suppl. \textbf{53}, 199-202 (1997)
 doi:10.1016/S0920-5632(96)00613-5
 [arXiv:hep-lat/9609023 [hep-lat]].
 A.~Hart \textit{et al.} [UKQCD],
 Nucl. Phys. B Proc. Suppl. \textbf{119}, 266-268 (2003)
 doi:10.1016/S0920-5632(03)01522-6
 [arXiv:hep-lat/0209063 [hep-lat]].
 T.~Burch, C.~Gattringer, L.~Y.~Glozman, C.~Hagen, C.~Lang and A.~Schafer,
 Phys. Rev. D \textbf{73}, 094505 (2006)
 doi:10.1103/PhysRevD.73.094505
 [arXiv:hep-lat/0601026 [hep-lat]].

\bibitem{Han:2013zg}
 H.~Y.~Han, X.~G.~Wu, H.~B.~Fu, Q.~L.~Zhang and T.~Zhong,
 Eur. Phys. J. A \textbf{49}, 78 (2013)
 doi:10.1140/epja/i2013-13078-7
 [arXiv:1301.3978 [hep-ph]].

\bibitem{Lees:2014iua}
 J.~Lees \textit{et al.} [BaBar],
 Phys. Rev. D \textbf{89}, no.11, 112004 (2014)
 doi:10.1103/PhysRevD.89.112004
 [arXiv:1403.7051 [hep-ex]].

\bibitem{Tanabashi:2018oca}
 M.~Tanabashi \textit{et al.} [Particle Data Group],
 Phys. Rev. D \textbf{98}, no.3, 030001 (2018)
 doi:10.1103/PhysRevD.98.030001

\bibitem{Bugg:2008ig}
 D.~Bugg,
 Phys. Rev. D \textbf{78}, 074023 (2008)
 doi:10.1103/PhysRevD.78.074023
 [arXiv:0808.2706 [hep-ex]].

\bibitem{Ablikim:2019hff}
 M.~Ablikim \textit{et al.} [BESIII],
 Chin. Phys. C \textbf{44} (2020) no.4, 040001
 doi:10.1088/1674-1137/44/4/040001
 [arXiv:1912.05983 [hep-ex]].

\bibitem{Asner:2008nq}
  D.~M.~Asner {\it et al.},
  Int.\ J.\ Mod.\ Phys.\ A {\bf 24}, S1 (2009)
  [arXiv:0809.1869 [hep-ex]].

\bibitem{Li:2016tlt}
  H.~B.~Li,
  Front.\ Phys.\ (Beijing) {\bf 12}, no. 5, 121301 (2017)
  doi:10.1007/s11467-017-0691-9
  [arXiv:1612.01775 [hep-ex]].

\bibitem{Bigi:2017eni}
  I.~I.~Bigi, X.~W.~Kang and H.~B.~Li,
  Chin.\ Phys.\ C {\bf 42}, no. 1, 013101 (2018)
  doi:10.1088/1674-1137/42/1/013101
  [arXiv:1704.04708 [hep-ph]].

\bibitem{Cheng:2013fba}
  H.~Y.~Cheng, C.~K.~Chua, K.~C.~Yang and Z.~Q.~Zhang,
  Phys.\ Rev.\ D {\bf 87}, no. 11, 114001 (2013)
  doi:10.1103/PhysRevD.87.114001
  [arXiv:1303.4403 [hep-ph]].

\bibitem{Ablikim:2018qzz}
 M.~Ablikim \textit{et al.} [BESIII],
 Phys. Rev. Lett. \textbf{122}, no.6, 062001 (2019)
 doi:10.1103/PhysRevLett.122.062001
 [arXiv:1809.06496 [hep-ex]].

\bibitem{Achasov:2018grq}
 N.~Achasov and A.~Kiselev,
 Phys. Rev. D \textbf{98}, no.9, 096009 (2018)
 doi:10.1103/PhysRevD.98.096009
 [arXiv:1805.10145 [hep-ph]].

\bibitem{Cheng:2018smm}
 X.~D.~Cheng, H.~B.~Li, R.~M.~Wang and M.~Z.~Yang,
 Phys. Rev. D \textbf{99}, no.1, 014024 (2019)
 doi:10.1103/PhysRevD.99.014024
 [arXiv:1812.00410 [hep-ph]].

\bibitem{Ablikim:2017auj}
  M.~Ablikim {\it et al.} [BESIII Collaboration],
  arXiv:1709.04323 [hep-ex].

\bibitem{Kornicer:2016axs}
  M.~Ablikim {\it et al.} [BESIII Collaboration],
  Phys.\ Rev.\ D {\bf 95}, no. 3, 032002 (2017)
  doi:10.1103/PhysRevD.95.032002
  [arXiv:1610.02479 [hep-ex]].

\bibitem{Ablikim:2018ffp}
  M.~Ablikim {\it et al.} [BESIII Collaboration],
  Phys.\ Rev.\ Lett.\  {\bf 121}, no. 8, 081802 (2018)
  doi:10.1103/PhysRevLett.121.081802
  [arXiv:1803.02166 [hep-ex]].

\bibitem{Buchoff:2008hh}
  M.~I.~Buchoff, J.~W.~Chen and A.~Walker-Loud,
  Phys.\ Rev.\ D {\bf 79}, 074503 (2009)
  doi:10.1103/PhysRevD.79.074503
  [arXiv:0810.2464 [hep-lat]].

\bibitem{Li:2009jd}
  H.~B.~Li,
  J.\ Phys.\ G {\bf 36}, 085009 (2009)
  doi:10.1088/0954-3899/36/8/085009
  [arXiv:0902.3032 [hep-ex]].

\bibitem{Ablikim:2012ur}
  M.~Ablikim {\it et al.} [BESIII Collaboration],
  Phys.\ Rev.\ D {\bf 86}, 092009 (2012)
  doi:10.1103/PhysRevD.86.092009
  [arXiv:1209.4963 [hep-ex]].

\bibitem{Ablikim:2014pfc}
  M.~Ablikim {\it et al.} [BESIII Collaboration],
  Phys.\ Rev.\ D {\bf 91}, no. 5, 052017 (2015)
  doi:10.1103/PhysRevD.91.052017
  [arXiv:1412.5258 [hep-ex]].

\bibitem{Ablikim:2016frj}
  M.~Ablikim {\it et al.} [BESIII Collaboration],
  Phys.\ Rev.\ Lett.\  {\bf 118}, no. 1, 012001 (2017)
  doi:10.1103/PhysRevLett.118.012001
  [arXiv:1606.03847 [hep-ex]].

\bibitem{Fang:2017qgz}
  S.~s.~Fang, A.~Kupsc and D.~h.~Wei,
  Chin.\ Phys.\ C {\bf 42}, no. 4, 042002 (2018)
  doi:10.1088/1674-1137/42/4/042002
  [arXiv:1710.05173 [hep-ex]].




\end{thebibliography}
\end{document}